\documentclass[conference]{IEEEtran}

\usepackage{colortbl}
\usepackage{longtable}
\definecolor{Gray}{gray}{0.85}
\usepackage[latin1]{inputenc}
\usepackage{amsmath}
\usepackage{amssymb}
\usepackage{graphicx}
\usepackage{xcolor}
\usepackage{enumerate}
\usepackage{todonotes}
\usepackage{supertabular,booktabs}
\usepackage{verbatim}
\usepackage{url}
\newcommand{\num}{13\space}

\newcommand{\libpath}{/Users/puix/pariya/research/library.bib}

\begin{document}
	\bstctlcite{IEEEexample:BSTcontrol} 
\title{Cross-Section Evidence-based Timelines for  Software Process Improvement Retrospectives:\\A Case Study of User eXperience Integration}

\author{
	\IEEEauthorblockN{Pariya Kashfi\IEEEauthorrefmark{1}, Robert Feldt\IEEEauthorrefmark{1}, Agneta Nilsson\IEEEauthorrefmark{1}, Richard Berntsson Svensson\IEEEauthorrefmark{2}}
	\IEEEauthorblockA{\IEEEauthorrefmark{1}Software Engineering Division\\
			Department of Computer Science and Engineering
			\\
			Chalmers University of Technology and Gothenburg University
		\\\{pariya.kashfi,robert.feldt,agnnil\}@chalmers.se}
	\IEEEauthorblockA{\IEEEauthorrefmark{2} Software Engineering Research Lab\\ School of Computing\\ Blekinge Institute of Technology
		\\\{rbs\}@bth.se}
}

\maketitle
\begin{abstract}
Although integrating UX practices into software development processes is a type of Software Process Improvement (SPI) activity, this has not yet been taken into account in UX publications.
In this study, we approach UX integration in a software development company in Sweden from a SPI perspective.
Following the guidelines in SPI literature, we performed a retrospective meeting at the company to reflect on their decade of SPI activities for enhancing UX integration.
The aim of the meeting was to reflect on, learn from, and coordinate various activities spanned across various organizational units and projects.
We therefore supported the meeting by a pre-generated timeline of the main activities in the organization that is different from common project retrospective meetings in SPI.  
This approach is a refinement of a similar approach that is used in Agile projects, and is shown to improve effectiveness of,  and decrease memory bias.
We hypothesized that this method can be useful in the context of UX integration, and in this broader scope.
To evaluate the method we gathered practitioners' view through a questionnaire.
The findings showed our hypothesis to be plausible.
Here, we present that UX integration research and practice can benefit from the SPI body of knowledge;
We also show that such cross-section evidence-based timeline retrospective meetings are useful for UX integration, and in a larger scale than one project, especially for identifying and reflecting on `organizational issues'.
This approach also provides a cross-section longitudinal overview of the SPI activities that cannot easily be gained in other common SPI learning approaches.

\end{abstract}								

\section{Introduction}
	To deliver a software that is consistent and of high quality, practitioners need to consider a large number of software quality characteristics~\cite{Chung2000a}. 
	A group of these quality characteristics relate to the development process and mainly concern developers (e.g. traceability), while another group are mainly important from the perspective of the end users (e.g. performance and usability)~\cite{Chung2000a}. 
	At an even broader level, the actual experience of the end users as they interact with the software needs to be taken into account.
	This widening scope of software quality characteristics has led to the introduction and study of the concept of user experience (UX). 
	Different definitions of UX exist but they share the same essence: 
	\textit{UX is a user's holistic perception of functionalities and quality characteristics of a software}~\cite{Hassenzahl2010a}.
	A good UX typically presumes that the software has high usability, but the latter does not automatically lead to the former~\cite{Hassenzahl2010a}.

	Research has shown that certain practices can increase the likelihood of delivering a positive UX~\cite{Hassenzahl2010a} (hereafter, \textit{UX practices}).
	But simply applying UX practices in isolation is not enough~\cite{Winter2010a}.
	Like methods and practices to support other software quality characteristics~\cite{Chung2000a} they need to be integrated into development processes and considered throughout projects (hereafter, \textit{UX integration}). 
	In other words, software development processes often need to be improved in order to better support these quality characteristics.
	This is generally referred to as  Software Process Improvement (SPI).
	Research shows that  different SPI methodologies (e.g. Capability Maturity Model (CMM)~\cite{Paulk1993a}, or ISO~9000-3~\cite{ISO9000-3}) have different impacts on various software quality characteristics~\cite{Ashrafi2003a}.
	For instance, Asharfi~\cite{Ashrafi2003a} empirically shows that ISO has a better impact on usability, while CMM on integrity, reliability, and testability.
	The impact of various SPI methodologies on UX of the software, however, remains an interesting question for researchers to address.

	Considering the lack of theories on the relation between SPI and UX of the product, it is expected to observe that often software companies approach UX integration in a more ad-hoc manner, without benefiting from known SPI methodologies, principles and guidelines, or aligning UX integration activities with other SPI activities in the company.
	A lack of SPI focus is also observable in studies on UX integration.
	The closest studies to this topic are those that directly or indirectly report on the relation between integrating usability practices into software development processes and SPI~\cite{Winter2010a,Oconnor2009}.
	In a recent mapping study on product quality within SPI activities, Garcia-Mireles et al.~\cite{Garcia-Mireles2015a} have identified two studies with usability in focus.
	First, a study by Winter et al.~\cite{Winter2010a} that compares eight years of experience in product related usability testing and evaluation with principles of SPI.
	The study is performed in an action research setting where the researchers take an active role in improving the process to better support usability testing.
	Secondly, a study by Oconnor et al.~\cite{Oconnor2009} that, through interviews with practitioners, investigates suitability of development processes for supporting usability.

	To the best of our knowledge, none of the current studies has investigated SPI principles and their role in success of UX integration. 
	Learning from past successes and failures in order to better plan and act on future improvements is one of the SPI principles~\cite{Dingsoyr2007a} that can be useful also in UX integration.
	One of the practical methods that is used for this purpose is `retrospective meetings' (aka. postmortem reviews) that make tacit experiences, explicit and enable these experiences to be better used in future~\cite{Dingsoyr2007a}.
	Retrospective meeting is also a simple and practical method for organizational learning \cite{Dingsoyr2007a}.
	To bridge the gap in UX integration body of knowledge, this paper investigates the usefulness of running retrospective meetings \textit{across organizational units and projects} (i.e. sections) to facilitate reflecting on, and coordinating UX integration activities.
	
	The rest of the paper is as follows:
	The first section gives a summary of UX in the context of software development, and a background on SPI and retrospective meetings.
	The second section describes the research method, and the third section presents the findings.
	In section four, we discuss the findings and present our reflection on usefulness of the proposed method.
	We end the paper with a conclusion and suggestion for future research.

\section{Background}
\label{sec:bg}
	As we elaborated in the introduction, an end user's perception of the functionalities and quality characteristics of a piece of software shapes her experience.
	It is therefore vital to consider UX of the software  in different stages of the development \textit{process}.
	 In requirements work for instance, UX needs to be taken into consideration when eliciting and prioritizing various functional and quality requirements~\cite{Ovad2015a}.
	One example would be prioritizing a new functionality that will most likely evoke a positive feeling (e.g. Gmail's reminder for forgotten attachment).
	Testing activities should also focus on not only functionalities and other quality characteristics, but also the UX: how will the end user perceive these functionalities and quality characteristics?
	Since taking UX of the software into account puts more emphasis on `the end users' needs', a trade-off between these needs and  `the business needs'\footnote{`Business' here refers to both software company's business in case of market-driven software, or customers' business in bespoke software.} becomes even more important, specially in case of conflicting needs~\cite{Winter2014a,Cajander2013a}. 
	One example is when the business requires having more banners on the website to earn money from advertisement while too much banners will bother the end users hence negatively affect their experience.
	Taking UX of software into account therefore asks for different sets of \textit{tools and methods} in order to support the above activities.
	It also requires new \textit{competences} to effectively perform different UX practices in various development stages.
	Therefore, efforts to enhance UX integration are types of SPI activities impacting  processes, practices, tools and methods, various stakeholders, structures, etc.
	
	SPI is shown to be an emergent rather than a deterministic activity~\cite{Allison2007a}.
	This highlights the importance of gaining an insight about how various activities have intertwined to inform each other and eventually influenced the SPI~\cite{Allison2007a}.
	This also indicates a possible divergence between intended and realized SPI activities and that SPI design and realization is shaped, i.e. enabled or constrained,  by its context~\cite{Allison2007a}.
	This puts more emphasis on understanding SPI from an organizational perspective, and looking at SPI changes from a holistic perspective:
	including both technical and organizational aspects~\cite{Allison2007a}.
	As research shows, organizational issues (e.g. organizational culture, and involved leadership) are at least as important in SPI as technical issues (e.g. models, practices, and tools)~\cite{Dyba2005b}.
	A number of these issues have been repeatedly brought up in research, including:
	 (i) effective communication and collaboration~\cite{Stelzer1999a,Zaharan1998a},
	 (ii) active staff involvement~\cite{Dyba2005b,Zaharan1998a,Goldenson1995a},
	 (iii) continuous learning and use of existing knowledge~\cite{Dyba2005b,Zaharan1998a}
	 (see~\cite{Dyba2005b} for an overview of organizational issues in SPI).
	
	\textit{Effective communication and collaboration} among various stakeholders and organizational units are shown to be two of the most influential factors in SPI success~\cite{Stelzer1999a,Zaharan1998a}.
	Stelzer~\cite{Stelzer1999a} emphasizes that an intensive communication and collaboration is needed in order to create an organizational culture that is in favor of improvements.
	Since often SPI activities are accompanied by misunderstandings, rumors, fears, and resistance from staff members, these issues need to be addressed through intensive communication~\cite{Stelzer1999a}.
	In addition, in any SPI activity, practitioners need to identify and reflect on  the challenges to any change in the organization as a result of SPI, in order to have an insightful plan for future improvements~\cite{Kezar2001}.
	\textit{Active staff involvement} is another main success factor in SPI~\cite{Stelzer1999a}, in particular involvement of technical staff and management~\cite{Dyba2001b}.
	If technical staff are not involved enough, as Goldenson and Herbsleb report~\cite{Goldenson1995a}, they can often feel that \textit{``SPI  gets in the way of their real work''} that in turn can negatively influence the SPI success.

	Therefore, SPI research highlights the importance of explicitly addressing such organizational issues~\cite{Dyba2005b}, in particular, through collectively learning from past experiences and making these experiences explicit~\cite{Dingsoyr2007a}.
	Dings\o{}yr~\cite{Dingsoyr2007a} emphasizes that such learning helps discovering points of improvement, or adjusting future SPI activities. 
	Performing project retrospectives therefore is  known to have a major positive impact on  SPI success~\cite{Rainer2002a}.
	On a higher level, a combination of different project retrospectives are known to provide insight to organizational learning~\cite{Dingsoyr2007a}.
	Considering that UX integration is also a type of SPI activity, it is reasonable to expect similar effects in case of UX integration as well.
	
	Nevertheless, to the best of our knowledge, the use of SPI principles, or practices, such as retrospective meetings, for UX integration remains unexplored. 
	Therefore, we designed a study to investigate the potential usefulness and benefits of retrospective meetings for reflecting on UX integration activities.	
	A decade of UX integration activities in the case company spanned across various organizational units and projects.
	So we found it important for practitioners to  get an overview of such activities and their interrelation since UX integration, similar to other SPI activities, is emergent, and shaped by intertwined activities in their context.
	In addition, the nature of UX,  being originated in non-technical fields, makes it even more important for technical practitioners to get a better understanding of what UX integration entails, and how that might affect their every day work.
	This implied that common single project retrospectives were a less suitable choice for this case, and therefore motivated running a retrospective meeting \textit{across sections}, covering the main projects running over the last decade, across various units in the case company.
	
	One known concern with retrospective meetings is that when a project has ended and project members get involved in other projects, they might quickly forget the details of the activities and their relations~\cite{Bjarnason2014}.
	This may result in biased discussions in the retrospective meeting.
	In our case, this problem was expected to be compounded considering the larger scope of the retrospective.
	To address memory bias in retrospective meetings, researchers~\cite{Bjarnason2014} suggest using Evidence-based timelines (EBT).
	This type of retrospective meeting is known as Evidence-based timelines Retrospective (EBTR).
	The idea behind EBTR is to use pre-constructed timelines in retrospective meetings instead of creating them during the meeting. 
	These timelines include various view points, that are time stamped and reflect different activities (e.g. decisions made, and events) performed over the course of a project.
	Bjarnason et al.~\cite{Bjarnason2014} argue that EBTs can prompt memory, save meeting time, provide objective information for the discussion, and more importantly minimize the risk of memory bias~\cite{Bjarnason2014}.
	EBTRs are described to be generic enough to be customized for different retrospective goals.
	Therefore, we hypothesized that this method can be useful in the context of UX integration, and to reflect on a broader scope of activities: spanned across various organizational units, projects and over a decade.

\section{Methods}
\label{sec:method}
	The case company is an international  software development company in Gothenburg, Sweden.
	It currently has around 2500 employees, and develops software for various domains, mainly aviation, using an agile development process. 
	The company is organized in three different units:
	(i) the first unit is responsible for developing the core features of the products,
	(ii) the second unit is responsible for customization of the features according to the needs of each customer,
	(iii) the third unit involves product owners and product managers and in large is responsible for strategic decisions and long term business planning.
	The company has recently hired a UX integration expert to take the responsibility of UX integration in the organization, and coordinate related activities, taking the role of a change agent.
	
	To prepare for and run the meeting, we  adapted the steps suggested by Bjarnason et al.~\cite{Bjarnason2014} for EBTRs in agile projects:
	
\subsection{Phase 1: Preparation} 
The first author was assigned as the meeting coordinator.
To create the timelines, she ran \num interviews in the company with practitioners from all the three units.
The interviewees were selected together with our main collaborators in the company, and based on these criteria: work experience in the company,  current and previous roles, and attitude towards UX integration (both positive and negative).
We also provided a written description of the interviewees we were looking for to prevent any misunderstandings regarding the criteria.
The interviews were face-to-face, audio recorded, and lasted between 30 to 60 minutes.
The interviews were transcribed and coded to identify the main UX integration activities.
Analyzing the interview data resulted in four main categories of activities\footnote{We use the generic term `activity' to refer to any type of decision, action, event, etc. that relates to UX integration, e.g. hiring a new UX integration expert is a type of UX integration activity, or changing the process from water-fall to agile is another activity that indirectly influences UX integration.}, to be visualized on the timelines.

\textbf{People:}
 People who have contributed to UX integration, including:
 (i) people who have been hired based on their UX-related competences, 
 (ii) have advocated UX integration, 
 or (iii) have been assigned UX-related roles. 
For each person, her name, competences, and specific UX-related responsibilities (if applicable) were presented on the people timeline.

\begin{figure}
	\centering
	\includegraphics[angle=0,scale=0.2]{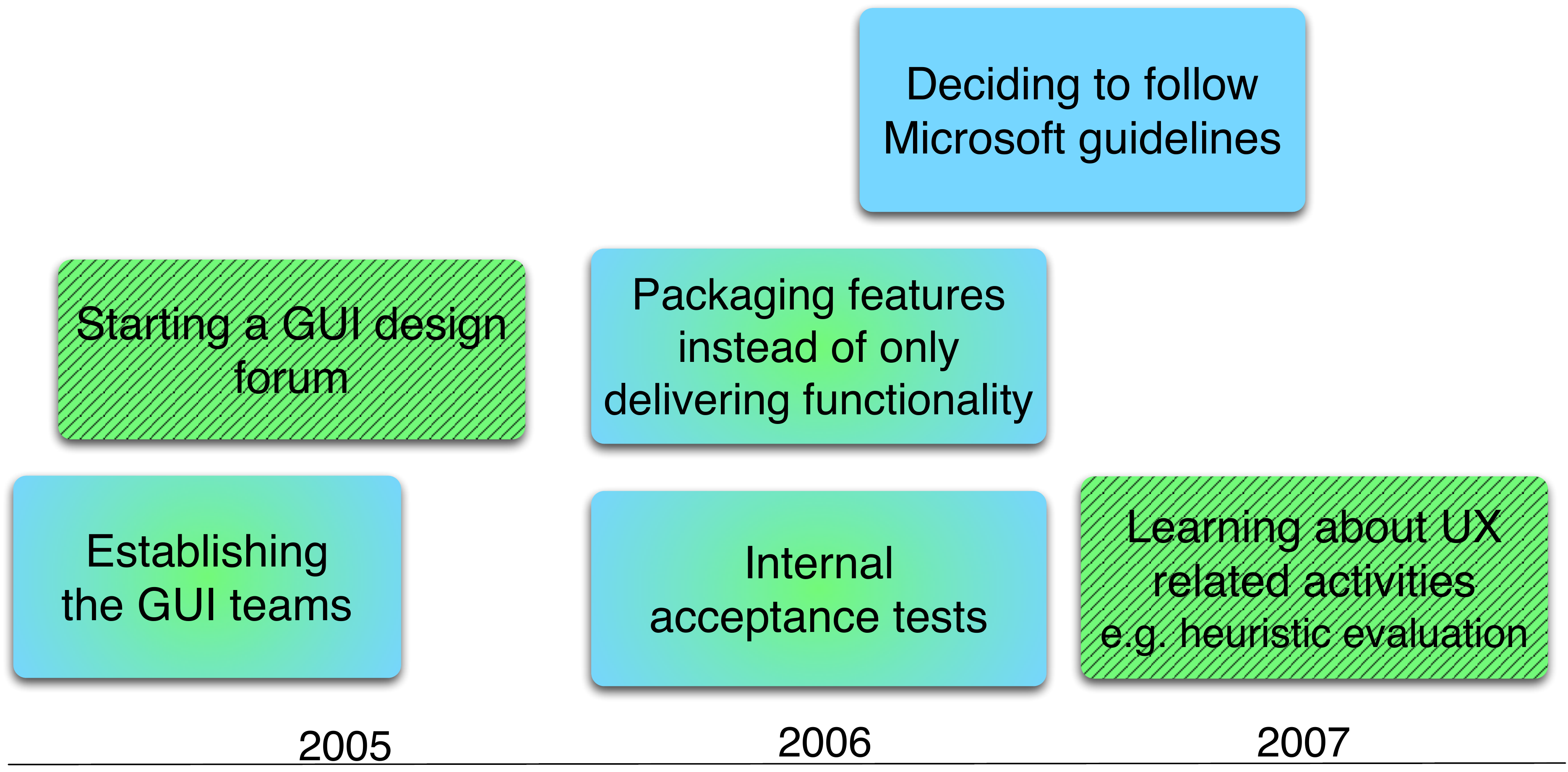}
	\caption{A part of the direct-events timeline. Each box represents one event, and is color coded to reflect in which unit of the company that has happened.}
		\label{fig:direct}
\end{figure}

\textbf{Direct-events:}
Activities that have been performed with an explicit aim to improve UX integration.
One example of a direct event is `starting study groups to learn about the concept of UX' (see Figure~\ref{fig:direct}).

\begin{figure}
	\centering
	\includegraphics[angle=0,scale=0.2]{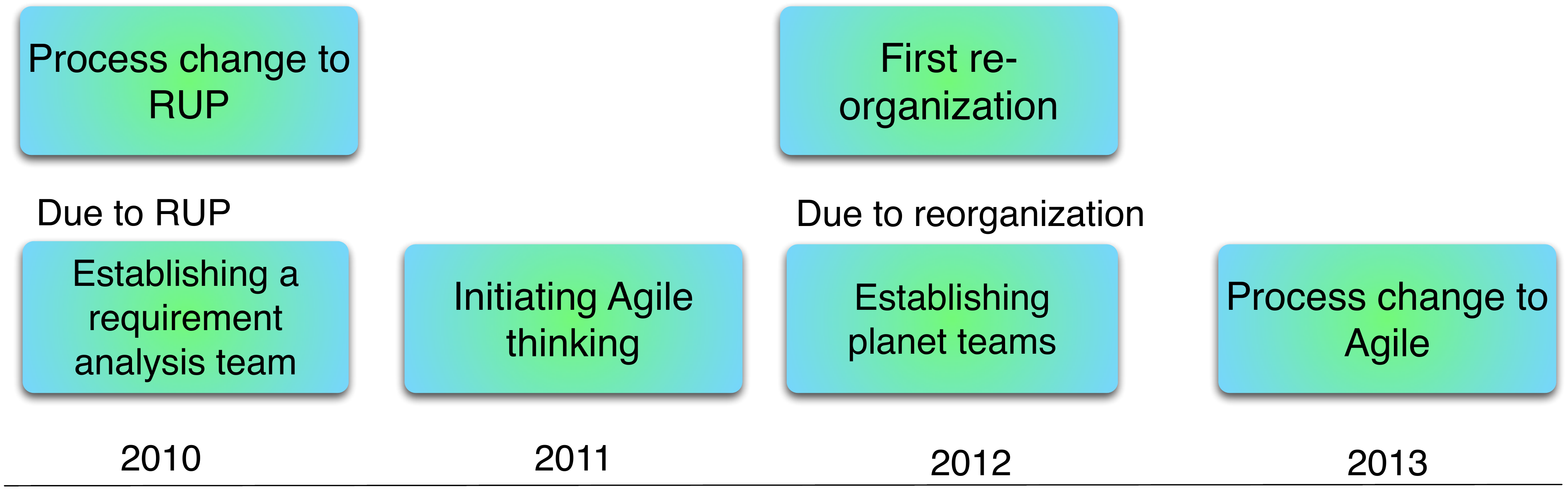}
	\caption{A part of the indirect-events timeline. Each box represents one event, and is color coded to reflect in which unit of the company that has happened.}
		\label{fig:indirect}
\end{figure}

\textbf{Indirect-events:}
Activities that have been performed for other purposes but indirectly influenced UX integration.
One example of an indirect event is `process change to agile' (see Figure~\ref{fig:indirect}).

\textbf{UX-artifacts:}
Tangible outputs of UX practices in  projects.
One example of a UX artifact is a `UX guideline'.

\subsection{Phase 2: Timeline Construction}
	 To increase readability of the timelines, for each of the above group of activities, one timeline was generated.
	 This was done in close collaboration with our main contacts in the company, and based on interview findings.
	When required, \textit{document analysis} was performed to discover more details about the activities.
	Where applicable, the timelines were color-coded to show in which unit of the company the activities have been performed.
	The distribution of the activities showed that `year' as the time unit on the timelines can be a suitable choice, and fits the goals of our  meeting: coordinating, and reflecting on the main activities, and their interrelation.

\subsection{Phase 3: Retrospective meeting}
The meeting was held at the case company and lasted 90 minutes.
Nine people were present in the meeting, representing the three units, at least two people from each unit.
 We should note that  the UX integration expert was not invited to the meeting for two reasons:
 (i) she is newly hired hence not informed about the history of UX integration in the company (i.e. the activities),
 (ii) part of the discussions in the meeting was going to be about her role and responsibilities and it was important that the participants could share their views freely.

\begin{figure}
  \centering
\includegraphics[angle=270,scale=0.3]{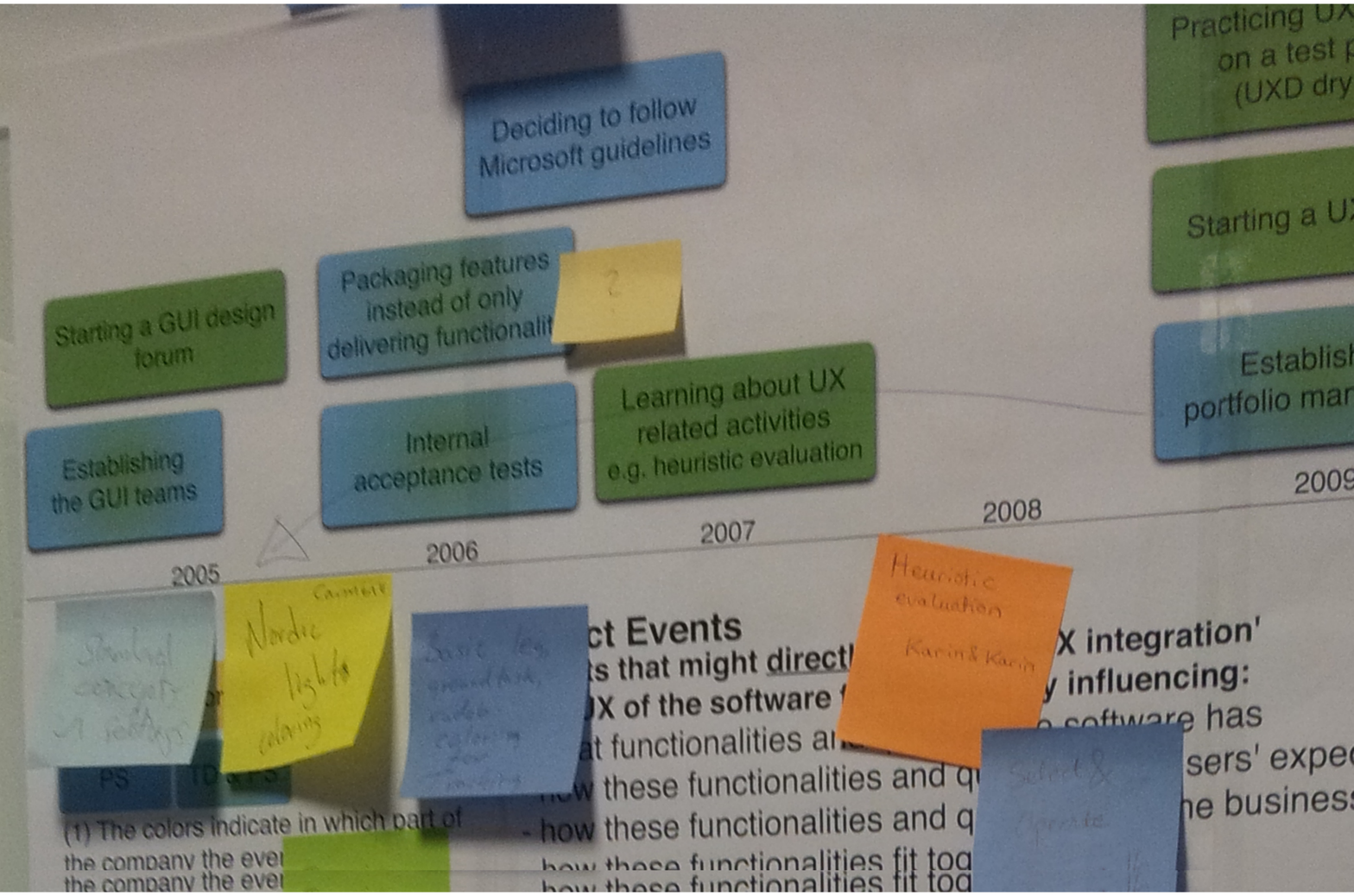}
\caption{A part of one of the timelines being annotated in the beginning of the \textit{cross-section EBT retrospective} meeting.
 As shown, to prevent misunderstandings definition of the terms used on the timelines was printed on each timeline.}
\label{fig:sample}
\end{figure}

We started the meeting by going through the aims of the meeting, and definitions of the terms used in the timelines.
Then the participants received 15 minutes to add their post-it notes (including additional, or updated data, or even questions regarding the activities) to the timelines (see Figure~\ref{fig:sample}).
To minimize the risks associated with including less details on the timelines (compared to the reported experiences with agile EBTRs~\cite{Bjarnason2014}), we invited the practitioners that have been employed for at least five years in the company, and have at least some knowledge regarding the presented activities.

The first and the last authors moderated the meeting, and took extensive notes to assure all the main discussion points are covered in the summaries.
The participants had open discussions about the timelines, and were guided by some focus questions if required.
The moderators tried to stay as neutral as possible in the discussions and let the participants freely exchange ideas and reflections.

\subsection{Phase 4: Reporting the findings of  the retrospective}
The summary of the meeting, and updated timelines were sent to the participants for validation, and getting approval. 
As a result, the timelines were finalized, and a retrospective report was prepared and distributed among the participants (i.e. representatives of various units), management, and the UX integration expert.
 
 \subsection{Validating The Cross-Section EBT Retrospective}
At the end of the meeting, a questionnaire was distributed among the participants to gather their thoughts about  the meeting and use of EBTs.
The questionnaire was inspired by the questionnaire included in the guidelines by Bjarnason et. al~\cite{Bjarnason2014}.
An example of a question included in the questionnaire is:
\textit{Through the retrospective meeting (including the timeline), to which degree did you gain new knowledge and insight about the big picture of the company UX integration activities?}\footnote{The questionnaires is available upon request.}
After initial analysis of the responses to the questionnaire, a number of follow up open-ended questions were sent to the participants to gather some additional data.

In order to investigate usefulness of the findings of the meeting for the UX integration expert, the results of the meeting (the retrospective report and the updated timelines) were shared with her,  with a series of open-ended questions to gather her reflection.
An example of such questions is:
\textit{How do you think such an approach (digging into history and for instance identifying these points of agreement and disagreement) can impact improving UX integration in the company?}


The analysis of our data is descriptive, and qualitative and emphasizes indication of the responses rather than their statistical significance, because the number of participants was too small for such an analysis.

\subsection{Threats to Validity}

Threats to validity are outlined and discussed based on the classification by Runeson and H\"{o}st~ \cite{Runeson2008a}.
To minimized the selection bias and threats to construct validity,  the subjects were selected together with our main contacts in the company, following a list of criteria we had prepared beforehand based on the goals of the meeting.
Also, to minimize the impact of presence of a researcher on the subjects' behavior and responses, confidentiality of the data was guaranteed in the interviews and meeting, and for the questionnaire responses.

To minimize threats to  internal validity, the interviews and the meeting were recorded either in audio or in form of extensive notes. 
The notes were shared with the participants for review.
In order to minimize the risk of misunderstanding among the meeting participants, the definition of the terms used on the  timelines were presented  in the beginning of the meeting.
A description of the method and the aim of the meeting was sent via email to the participants two days before the meeting.
The timelines were also printed in a smaller size and distributed among the participants one day before the meeting to give them a chance get prepared for the meeting.
\\ 
Regarding external validity, the method has been validated in other cases~\cite{Bjarnason2014} that can help in comparing how EBTs can be used in different settings.
Nevertheless, running such a meeting in this broad context and across-sections is based on a single case study and its findings should be used by caution.
The study is inherently qualitative and does not attempt to generalize beyond the actual setting, and is more concerned with explaining and understanding the phenomena under study.

\section{Results and Analysis}
\label{sec:results}

	During the meeting, the participants repeatedly pointed to the timelines to refer to different activities (i.e. events, people or artifacts).
	They also discussed the interrelation among a number of the activities, in particular between the UX artifact timeline and the direct-event timeline.
	We also observed how the printed timelines initiated discussions about details of the activities and how the participants' perceived positive or negative impact of these activities on UX integration in the company.
	Here, we report on the experiences from using the method, based on our observations in the meeting, the participants' responses to the questionnaire, and reflection of the company's UX integration expert and management.
	
	According to the questionnaire responses, the participants generally found this method useful for learning about and reflecting on UX integration in the company.
	They emphasized that the method provided a `good overview' and a `big picture' of the UX integration activities, and that this overview made it easier to  reflect on the activities.
	They also highlighted that the timeline helped them get informed about the activities performed in different units.
	This was not clear to all participants before the meeting in particular since these units lack communication (at least) concerning UX integration.
		
	In addition, the meeting facilitated communication among the participants according to them.
	Regarding this one participant stated:
	\textit{``I think this is a very good technique. It keeps everyone focused, one can come prepared, and you get the chance to steer the meeting to what you believe is important. Of course there is always the risk to steer too much and miss out on some of the `out-of-the-box' thinking.''}
	Another participant stressed the usefulness of the timelines as a `base for the discussions':
	\textit{``The timelines helped us to get a common understanding about what had happened during the years, and we could point on the timelines and have them as a base for the discussion.''}
	According to the participants, the timelines provided `a neutral' and `factual' basis to reflect on the activities.
	Also, one participant highlighted that the retrospective meeting was much `calmer' than other informal discussion meetings  about UX integration because of this neutral and factual basis, and presence of  `a neutral moderator'.
	Especially since current conflicts among various units, or even individuals concerning UX integration often prevents 'constructive discussions' about it.
	Emphasizing this, one participant stated:
	 \textit{``I liked that people from all three departments sat together to share some of their views. This helps us align for future actions''}
	 	 
	In the meeting, we observed that the `direct-event', and `UX artifact' timelines opened up a series of discussions about different UX integration activities that have mostly been bottom-up in some teams and based on individual interest of developers.
	Through the people timeline, it also became obvious to the participants that some  of these activities have been performed by people with no UX-related role or responsibility.
	Regarding the connections, one practitioner highlighted:
	\textit{``I liked the overview and the visibility of when things happened in relation to each other.''}
	The participants' view on this matter was however divided.
	While some participants believed they were able to identify some connections, the others expressed that it was not easy to identify them.
		
	The participants also highlighted that the meeting facilitated identifying, and making the main points of agreement and disagreement concerning UX integration explicit.
	For instance, it became clear to the participants that most activities have been 'ad-hoc', with not enough leadership, support from management, or even mandate. 
	It became also clear that these activities have not been coordinated (e.g. similar activities duplicated in different units without alignment, or artifacts being introduced without informing roles that would receive these artifacts in later stages of the project).
	As one of the participants pointed out: \textit{``the difference between people and the lack of formal work became obvious.''}.
	Another participant responded that the meeting helped them to agree on main activities that have been performed in the past and influenced UX integration.
	Similarly, another participant stated: \textit{``It is clear that we don't agree on the UX methods and processes and it is good that is out in the open now.''}
	In addition,  the participants generally agreed that the meeting was helpful in emphasizing this limited communication concerning UX integration.
	For instance, one participant highlighted that the  meeting facilitated `communication' between different roles with different perspectives, and further emphasized: 
	\textit{``I think It was clear why we don't have a common understanding about how and what to do with UX: [limited] communication.''}
	
	We observed that the meeting discussions about the activities, and in particular learning about different viewpoints helped the participants generate ideas for future improvements.
	For instance, one participant stated: \textit{``We need one person that is in charge [of UX integration]  \dots I think we need to look at which kind of profile we need for these tasks.''}
	Another participant stated: \textit{``I think I got a good understanding of what needs to be changed.''}
	
	The UX integration expert was very positive towards the findings of the meeting, and using them to design future UX integration improvement activities for the company.
	She emphasized:
	\textit{``Based on the information from the meeting I think I have a better
	understanding of where they are coming from ... I have
	to show them what use they can get from a UX-person.''}
	She found the findings providing useful and valuable insight for her to understand the challenges with UX integration over the years and how they have influenced the attitude of practitioners towards UX in general.
	She stressed: \textit{``I think this kind of information is very useful to know for a new UX person coming to the company.''}
	She further emphasized, such information is difficult to gather in particular for practitioners like her that are new to an organization.
	One example of such information is how over the years various people have influenced UX integration (i.e. the people timeline) and how this has been perceived by other practitioners (i.e. the discussions around the people timeline in the meeting).
	Another interesting use of the findings according to her was learning about the `expectations' for her role as  `UX integration expert', and the people's attitude towards UX in general.
	Regarding this, she stated:
	\textit{``They expected a totally different approach in their UX work than the one I am offering to them right now.''}
	The expert argued for another potential interesting use of \textit{cross-section EBT retrospective} meetings stating:
	\textit{``this kind of meeting could be fruitful to do as a UX-consulting company together with presumptive clients, as a free selling activity for UX. The consults could visualize how the clients company have treated UX over the years and come with possible solutions on how to
		work better with it.''}
	
	In addition, in a separate meeting, the meeting findings were presented to the senior management group in the case company.
	The management appreciated getting an overall image of the company's UX integration activities.
	They described the result of the retrospective meeting as: ` constructive', `interesting' and `thought-provoking'.
	They also highlighted that they can use this information to learn from past experiences and accordingly adjust future UX integration improvement activities.
		
	A number of open questions were included in the questionnaire to investigate how the retrospective meeting or the timelines could be improved, e.g. ``What additional data would be beneficial to include in the timelines?'' 
	One of the participant suggested including actual decisions that have affected UX of the products.
	Another participant suggested showing a history of some of the UX guideline documents in the company and how they have evolved over time.
	Two other participants stated that they would prefer to have  longer timelines including the activities performed even before 2005.
	
	As we showed,  that this method can facilitate a longitudinal analysis of UX integration (or other SPI activities) however this is a bonus rather than the main focus of the method.
	The main focus of the method which we emphasize in this study is providing the \textit{cross-section} overview of the activities, and their relation to each other and to UX integration improvement.
	Nevertheless, because of the longitudinal nature of the timelines, we applied theories on \textit{longitudinal research}~\cite{Street2012} when creating them.
	Below, based on these theories, and our experience in applying the method, we present a guideline for generating the timelines, in particular for selecting their \textit{time unit}, \textit{time boundaries}, and the \textit{time period} that according to these theories are important considerations in longitudinal studies~\cite{Street2012}.
	The guideline is tabulated in Table~\ref{tbl:guidleine}.
	These guidelines can be used together with the timelines for EBTRs provided by Bjarnason et al.~\cite{Bjarnason2014} to guide practitioners in preparing for and running the \textit{cross-section EBT retrospective} meeting.
	
	\textit{Time unit} refers to the granularity of the timeline (for instance in our case we presented the activities yearly).
	Time unit depends on the frequency of the improvement activities performed in the company, the higher the frequency the smaller the time unit.
	Regarding the details included in the timelines, we should emphasize that extracting details about all activities (although wished by some of the participants) may not be useful or even possible.
	True influence of each activity on UX integration can only be identified based on how much UX of the products are influenced at the end, which is not an easy question to answer, and the efforts to find an answer may not be justified.
	Therefore, the details should be kept at a level providing enough information for the participants to understand and reflect on the activities and how they intertwined.
	This also naturally depends on previous knowledge of the participants about these activities.
	
	\textit{Time boundary} refers to 'the window in time' during which we want to study the activities (for instance we included the activities from 2005 to present, in our case 2015).
	This is not relevant for creating EBTs in project retrospectives (i.e. EBTRs) because the time frame in project EBTRs is fixed, i.e. the project start and end.
	For \textit{cross-section EBT retrospective} meetings, however, this depends on when the improvement activities have been started.

	\textit{Time period} refers to where the window in time (i.e. the time boundary) is located in the whole history of the company (for instance if one decides to study two years of activities, should this two years be 2012-2014 or 2013-2015?).
	Similarly, 	this is not relevant for project EBTRs because the time frame is fixed, and often the project in such a meeting is not analyzed in timely relation to other activities in the company.
	For \textit{cross-section EBT retrospective} meetings the time period should often include 'present', unless the aim of the meeting is specifically to reflect on a certain group of activities in a period of time, for instance to study how they have intertwined.
	\begin{table}
		\begin{footnotesize}

			\begin{tabular}
				{|p{0.09\textwidth}| p{0.35\textwidth}|}
				
				\hline 
				\textbf{Consideration}    & \textbf{Tips}  \\ 
				\hline 
				Time unit

				& Justify the time unit based on the the pace of UX integration activities: the more frequent the activities the smaller the unit
				\\ \hline
				Time boundaries  
						& Justify the boundaries of the timeline based on when the improvement activities have been started 
				\\ \hline
				Time period 
						& The period often includes 'present', unless the aim of the meeting is specifically to reflect on a certain group of activities performed over the time period
								\\ \hline
				Meeting length
						& To be able to sufficiency discuss and reflect on all activities and their relation, hold at least a 4 hour meeting.
				\\ \hline
				
				Participants  
						& UX integration impacts both UX and non-UX professionals, so representatives of both groups should participate in the meeting.
				\\ \hline
				Preparation
				
				& If the coordinator is internal with more knowledge on previous activities in the company, less interviews are needed to identify the key activities.
				Level of details provided based on the interview findings and studying the evidence also depends on the knowledge of the participants about the included activities.
				\\ \hline
				
			\end{tabular}
		\end{footnotesize}
		\caption{Guidelines for generating the cross-section timelines}
		\label{tbl:guidleine}
	\end{table}

	\section{Discussion}
	\label{sec:disc}
	Our hypothesis that a \textit{cross-section EBT retrospective} meeting could be useful for reflecting on, learning from, and coordinating UX integration activities is plausible based on our findings.
	According to the questionnaire responses, the participants generally agreed that the meeting helped them to (i) have a factual discussion in the meeting, 
	(ii) remember and agree on UX integration activities, and activities that indirectly influenced UX integration 
	(iii) remember details of those activities to a good extent, 
	(iv) get a big picture and an overview of the decade of activities.
	These findings are in line with previous studies on EBTRs in context of agile projects~\cite{Bjarnason2014}.
	The participants, however, had a divided view on whether the meeting (and in particular the EBT) provided sufficient data to identify connections between the activities.
	Still in previous uses of EBTRs, this has been reported as one of the benefits of EBTs.
	Below we discuss the relevance of our findings in the context of UX integration and SPI in general.
	
	\textit{Providing an overview of the activities:} 
	Practitioners' 	statements that this approach gave them a `good overview of the activities' in particular `activities performed in different units' is evidence for usefulness of the method in providing a cross-section overview of the activities across projects and organizational units.
	Such an overview is not likely to be gained through separate project retrospectives that is a more common learning practice in SPI. 
	This is because project retrospectives focus on a certain project and cannot directly provide \textit{a cross-section} overview of the improvement activities.
	As emphasized by the UX integration expert, one use of such an overview is to find out  `how the company has so far worked with UX integration'.
	This  provides more insight about the company's history that is useful for planning future improvements as the expert highlighted as well.
	Another use of such an overview, as we explain below, is to help understand how these cross-section activities are intertwined.
	
	\textit{Visualizing connection between the activities}:
	An overview of the activities can facilitate gaining insight about how these various activities have affected each other.
	We showed that the meeting participants to some extent were able to identify the connection between various activities on a timeline or across timelines.
	This is evidenced by practitioners' highlighting one use of the timelines as  `visibility of activities in relation to each other'.
	Insight about the intertwined activities, as we discussed in Section~\ref{sec:bg}, is an important input to analysis of SPI activities to increase its success.
	However, considering the divided view of practitioners,  this remains open for further investigation how to better support identifying these connections.
	One possible solution for instance could be that the moderators better support the participants in identifying these connections by asking specific questions that draws attention to potential connections among various timelines or activities.
	
	\textit{Facilitating communication:}
	Our experience shows that the \textit{cross-section EBT retrospective} meeting can facilitate practitioners' communication about UX integration through making points of agreement and disagreement explicit, helping in identifying a number of the main challenges to UX integration, and providing an overview of the improvement activities.
	Despite the importance of communication concerning UX integration, the concept of communication is often highlighted in the UX literature with emphasis on communicating UX goals, artifacts and communication between UX and non-UX practitioners in the projects (e.g.~\cite{Cajander2013a}).
	A SPI perspective, however, emphasizes communicating how these goals, artifacts or roles are going to be introduced to the current working process in a company.
	
	Making points of agreement and disagreement explicit was based on our experience one benefit of the \textit{cross-section EBT retrospective} that can be in particular useful in case of conflicts in the organizations (as was the case in this company).
	The participants for instance appreciated that some of the main points of disagreement concerning UX integration activities `became obvious' through the meeting, and were `out in the open'.
	Because this approach, to a good extent, prevents biased discussions using factual and unbiased timelines, as also emphasized by the participants.
	They for instance stressed that the meeting was `calmer' than expected considering the existing conflicts among various units.
	The UX integration expert found the results of the meeting informative, among others, concerning `expectations for her role'.
	 In her view, such an information is valuable for a `new person coming in' to any organization to take the role of a change agent to improve UX integration.
	She stated this information had not been explicitly communicated to her before the meeting.
		
	\textit{Identifying improvements for future:}	
	Our experience also shows that the meeting facilitates generating ideas for future improvements and hels practitioners to collectively identify `what needs to be changed' as expressed by one of the participants.
	As another participant emphasized, the discussions in the meeting can help them better `align for future'.
	This directly relates to our previous discussion about how the meeting facilitated communication, for instance by helping the participants identify main points of agreement and disagreement and challenges to UX integration.
	The UX integration expert also stated that the information gained from the meeting provided ideas for `possible solutions' considering the history of the UX integration activities; it also inspired ideas for `how to work better with UX integration' considering such a context.
		
	\textit{Facilitating active staff involvement:}
	One of our observations in this study was that the \textit{cross-section EBT retrospective}  meeting provided a venue for practitioners with different backgrounds to actively get involved in this reflection and learning activity. 
	Through this meeting, both technical and business staff got a chance to raise their concerns about UX integration and how that could impact their everyday work or short-term or long-term plans for the projects at hand.
	There is however no guarantee that such an involvement will be pursued throughout future UX integration improvement activities,  but the meeting is at least a starting point for that.
	We hypothesize that if performed more frequently, such a meeting can be a `milestone' to assure active staff involvements throughout the activities.
	This remains an open questions for research to investigate.
	
	The topics we discussed above belong to the \textit{organizational aspects} of UX integration (or SPI in general).
	Therefore, we argue that this method can support practitioners in dealing with organizational issues.
	This is in particular important since despite their importance, the organizational issues related to UX integration still remain less explored~\cite{Winter2014a}.
	The findings of this study however should be used with caution considering that they are based on a single case study.
	We plan to perform more empirical investigation of the method, specially since a number of other companies  have shown interest in using this methods.
	Currently, we are planning at least one other case study of the method in a large Telecom company in Sweden.
	This study is an evidence that UX integration practice can benefit from the SPI body of knowledge considering the similarities between the two topics.
	We therefore encourage other UX practitioners and researchers to also take a SPI perspective, when applicable,  to benefit more from SPI guidelines, principles and methods in thier efforts to improve UX integration in software companies.
		
	In addition, considering the similarities of UX integration activities and other SPI activities, it is reasonable to assume that the benefits of \textit{cross-section EBT retrospective} meetings experienced in this case study can be observed for other SPI activities as well.
	We therefore suggest using this method by software development practitioners in order to reflect on, learn from and coordinate other types of SPI activities.
	This method adds to the practitioners' toolset of methods for supporting a successful SPI.
	Such a retrospective meeting can be useful in particular since research shows that still analyzing the outcome of several reviews to facilitate learning on an organizational level is not a common practice in software companies~\cite{Dingsoyr2007a}.
	This admittedly  is  a subject that needs to be investigated more in future, for instance through other empirical studies that use the method to coordinate and reflect on other SPI activities.

\bibliographystyle{IEEEtran}
\bibliography{nourl,\libpath}

\end{document}